\documentclass[]{elsarticle} 
\usepackage[hyphens]{url}

  \journal{An Elsevier Journal} 

\usepackage{lineno} 
\providecommand{\tightlist}{%
  \setlength{\itemsep}{0pt}\setlength{\parskip}{0pt}}

\biboptions{sort&compress} 
\usepackage{graphicx}
\usepackage{booktabs} 

\usepackage[T1]{fontenc}
\usepackage{lmodern}
\usepackage{amssymb,amsmath}
\usepackage{ifxetex,ifluatex}
\usepackage{fixltx2e} 
\IfFileExists{upquote.sty}{\usepackage{upquote}}{}
\ifnum 0\ifxetex 1\fi\ifluatex 1\fi=0 
  \usepackage[utf8]{inputenc}
\else 
  \usepackage{fontspec}
  \ifxetex
    \usepackage{xltxtra,xunicode}
  \fi
  \defaultfontfeatures{Mapping=tex-text,Scale=MatchLowercase}
  
\fi
\IfFileExists{microtype.sty}{\usepackage{microtype}}{}
\usepackage{longtable}
\usepackage{graphicx}
\makeatletter
\def\maxwidth{\ifdim\Gin@nat@width>\linewidth\linewidth
\else\Gin@nat@width\fi}
\makeatother
\let\Oldincludegraphics\includegraphics
\renewcommand{\includegraphics}[1]{\Oldincludegraphics[width=\maxwidth]{#1}}
\ifxetex
  \usepackage[setpagesize=false, 
              unicode=false, 
              xetex]{hyperref}
\else
  \usepackage[unicode=true]{hyperref}
\fi
\hypersetup{breaklinks=true,
            bookmarks=true,
            pdfauthor={},
            pdftitle={A Universal Algorithm for Continuous Time Random Walks Limit Distributions},
            colorlinks=true,
            urlcolor=blue,
            linkcolor=magenta,
            pdfborder={0 0 0}}
\usepackage{amsthm,amsmath,amssymb,hyperref}

\begin{document}
\begin{frontmatter}

  \title{A Universal Algorithm for Continuous Time Random Walks Limit
Distributions}
    \author[UNSW Sydney]{Gurtek Gill}
   \ead{rickygill01@gmail.com} 
  
    \author[UNSW Sydney]{Peter Straka\corref{c1}}
   \ead{p.straka@unsw.edu.au} 
   \cortext[c1]{Corresponding Author}
      \address[UNSW Sydney]{School of Mathematics \& Statistics, Sydney, NSW 2052, Australia}
  
  \begin{abstract}
  In this article, we generalize the recent Discrete Time Random Walk
  (DTRW) algorithm, which was introduced for the computation of
  probability densities of fractional diffusion. Although it has the same
  computational complexity and shares the same desirable features
  (consistency, conservation of mass, strictly non-negative solutions), it
  applies to virtually every conceivable Continuous Time Random Walk
  (CTRW) limit process, which we define broadly as the limit of a sequence
  of jump processes with renewals at every jump. Our only restrictive
  assumption is the boundedness and continuity of coefficients of the
  underlying Langevin proceesses.
  
  We highlight three main novel use-cases: i) CTRWs with spatially varying
  waiting times, e.g.~for interface problems between two differently
  anomalous media; ii) (varying) temporal drift, which limits the
  short-time speed of subdiffusive processes; and iii) the computation of
  probability densities for generalized inverse subordinators.
  \end{abstract}
 \end{frontmatter}

\newtheorem{lemma}{Lemma}

\section{Introduction}\label{introduction}

Subdiffusive transport processes are characterized via a sublinear
growth of the mean squared displacement:
\(\langle X_t \rangle \sim t^\alpha\), where \(0 < \alpha < 1\). Such
processes are usually modelled either by fractional Brownian motion or
Continuous Time Random Walks (CTRWs), depending on whether the
auto-correlation of jumps decays slowly or the waiting times between
jumps are heavy-tailed with parameter \(\alpha\), modelling traps or
dead ends (Henry et al. 2010). The CTRW model has proven to be a
particularly useful model, predominantly in biophysics (Metzler and
Klafter 2000; Tolić-Nørrelykke et al. 2004; Wong et al. 2004; Banks and
Fradin 2005; Santamaria et al. 2006; Höfling, Franosch, and Article
2012; Regner et al. 2013), but also in groundwater hydrology (Berkowitz,
Emmanuel, and Scher 2008; Schumer et al. 2003) and econophysics (Scalas
2006).

A modelling framework for the evolution of probability densities of
random walks is given by the Fokker--Planck equation (Gardiner 2004):

\begin{align}
\frac{\partial P(y,t)}{\partial t} = \mathcal L^*(y,t) P(y,t) + \delta_{(0,0)}(y,t),
\end{align}

where

\begin{align}
\mathcal L^* g(y,t)
&= -\frac{\partial }{\partial y}[b(y,t) g(y,t)]
+\frac{1}{2}\frac{\partial^2 }{\partial y^2}[a(y,t) g(y,t)]
\end{align}

is called the Fokker--Planck operator. CTRWs generalize random walks by
allowing a larger, heavy-tailed class of waiting times before each jump.
This translates into a \emph{memory kernel} \(V(y,t)\) acting on the
time variable in the equation (Baeumer and Straka 2016):

\begin{align}
\frac{\partial P(y,t)}{\partial t} = \mathcal L^*(y,t) \left[ \frac{\partial}{\partial t}
\int_0^t P(y,t-s) V(y,s)\,ds \right] + \delta_{(0,0)}(y,t).
\label{eq:FFPE}
\end{align}

The table below gives an overview over frequently studied forms of
\(V(y,s)\):

\begin{longtable}[]{@{}rccc@{}}
\toprule
\begin{minipage}[b]{0.10\columnwidth}\raggedleft\strut
\strut
\end{minipage} & \begin{minipage}[b]{0.20\columnwidth}\centering\strut
Kernel\strut
\end{minipage} & \begin{minipage}[b]{0.20\columnwidth}\centering\strut
Laplace Transform\strut
\end{minipage} & \begin{minipage}[b]{0.21\columnwidth}\centering\strut
Reference\strut
\end{minipage}\tabularnewline
\midrule
\endhead
\begin{minipage}[t]{0.10\columnwidth}\raggedleft\strut
no memory\strut
\end{minipage} & \begin{minipage}[t]{0.20\columnwidth}\centering\strut
\(1\)\strut
\end{minipage} & \begin{minipage}[t]{0.20\columnwidth}\centering\strut
\(\lambda^{-1}\)\strut
\end{minipage} & \begin{minipage}[t]{0.21\columnwidth}\centering\strut
\strut
\end{minipage}\tabularnewline
\begin{minipage}[t]{0.10\columnwidth}\raggedleft\strut
subdiffusion\strut
\end{minipage} & \begin{minipage}[t]{0.20\columnwidth}\centering\strut
\(s^{\alpha-1} / \Gamma(\alpha)\)\strut
\end{minipage} & \begin{minipage}[t]{0.20\columnwidth}\centering\strut
\(\lambda^{-\alpha}\)\strut
\end{minipage} & \begin{minipage}[t]{0.21\columnwidth}\centering\strut
Sokolov and Klafter (2006)\strut
\end{minipage}\tabularnewline
\begin{minipage}[t]{0.10\columnwidth}\raggedleft\strut
tempered subdiffusion\strut
\end{minipage} & \begin{minipage}[t]{0.20\columnwidth}\centering\strut
unknown\strut
\end{minipage} & \begin{minipage}[t]{0.20\columnwidth}\centering\strut
\(((\lambda + \theta)^\alpha - \theta^\alpha)^{-1}\)\strut
\end{minipage} & \begin{minipage}[t]{0.21\columnwidth}\centering\strut
Gajda and Magdziarz (2010)\strut
\end{minipage}\tabularnewline
\bottomrule
\end{longtable}

As the table indicates, most researchers have studied spatially constant
memory kernels, without any dependence on the space variable \(y\). This
implies a homogeneous distribution of waiting times throughout the
entire medium, i.e.~that diffusion is equally anomalous everywhere. This
assumption is of course too restrictive for some applications in
biophysics (Wong et al. 2004; Straka and Fedotov 2015), e.g.~when
trapping varies due to locally different compositions of the cellular
matrix. Moreover, media with two different anomalous exponents exhibit
interesting, paradoxical behaviour (Korabel and Barkai 2010; Straka
2018), and have been studied (analytically) in the physics literature
(Stickler and Schachinger 2011; Fedotov and Falconer 2012).

Numerous methods for the computation of solutions to \emph{homogeneously
anomalous} diffusion have been developed, among them explicit methods
(Yuste and Acedo 2005), implicit methods (Langlands and Henry 2005),
spectral methods (Li and Xu 2009; Hanert and Piret 2014) and Galerkin
methods (Mustapha and McLean 2011). In the domain of
\emph{inhomogeneously anomalous} diffusion, several authors have
developed computational methods for \emph{variable order} fractional
Fokker--Planck Equations, but only the equation studied by Chen et al.
(2010) is consistent with a CTRW scaling limit representation (Straka
2018).

The algorithm we introduce in this paper is an extension of the
Semi-Markov approach by Gill and Straka (2016). It computes solutions to
all Fokker--Planck equations of type \eqref{eq:FFPE} with spatially
varying memory. Its only requirement is that the coefficients of the
underlying bivariate Langevin process \((Y_u, Z_u)\), which tracks the
location resp. current time, are bounded and continuous and can be
evaluated numerically.

Similarly to the Discrete Time Random Walk (DTRW) method (C. N.
Angstmann, Donnelly, Henry, and Nichols 2015; Angstmann et al. 2016),
our algorithm calculates the probability distributions of a CTRW whose
waiting times are grid-valued, and which approximates the continuum
limit process. The advantages of this approach are that mass is
necessarily conserved in each timestep; that solutions are guaranteed to
be nowhere negative; and that stochastic process convergence implies the
consistency of the algorithm. However, we do not rely on discrete
Z-transforms, which means that our method remains tractable not just for
Shibuya-distributed waiting times.

This paper is organized as follows:

\begin{description}
\tightlist
\item[Section 2:]
We give a short account of bivariate Langevin dynamics characterizing
CTRW limit processes.
\item[Section 3:]
We construct a sequence of DTRWs which converges to a CTRW continuum
limit process, represented by a general bivariate Langevin equation
\((Y_u, Z_u)\).
\item[Section 4:]
We calculate the probability distributions of the DTRW via genearlized
master equations in an extended tate space.
\item[Section 5:]
We study three novel use-cases, namely an interface problem, spatially
varying temporal drift, and inverse subordinators.
\item[Section 6:]
concludes.
\end{description}

\section{Stochastic solution to Fokker-Planck equation with
memory}\label{stochastic-solution-to-fokker-planck-equation-with-memory}

The Langevin representation of a stochastic process whose distribution
\(P(y,t)\) solves a Fokker-Planck equation with memory has been studied
in various articles (Weron and Magdziarz 2008; Henry, Langlands, and
Straka 2010; Gajda and Magdziarz 2010; Hahn et al. 2011). Recently, a
Langevin representation for inhomogeneous anomalous diffusion was given
(Straka 2018): Consider the bivariate Langevin process with state space
\(\mathbb R \times [0,\infty)\)

\begin{align}
dY_u &= b(Y_u, Z_u)\, du + \sqrt{a(Y_u, Z_u)} \, dW_u
\label{eq:dY}\\
dZ_u &= d(Y_u) \, du + \int_{w > 0} w\, n(dw, du)
\label{eq:dZ}
\end{align}

Here, \(u\) is auxiliary time, corresponding to the number of jumps;
\(b(y,t)\) and \(a(y,t)\) are drift and diffusivity coefficients (of
units length resp. length\(^2\) per unit \emph{auxiliary time})
appearing in \eqref{eq:FFPE}; \(d(y)\) is a temporal drift coefficient
(unit physical time per unit auxiliary time). Finally, \(n(dw, du)\)
denotes Levy noise that can be spatially varying. Recall that Levy noise
has a representation as a Counting Measure, where for any rectangle
\(R = (u_1, u_2) \times (w_1, w_2) \subset [0,\infty) \times (0,\infty)\)
the number of points \(n(R)\) in \(R\) is Poisson distributed, and
independent of any counts in other, disjoint rectangles (Applebaum
2009). The Poisson distribution, and hence the entire Counting Measure,
is governed by a unique mean measure \(m(dw, du)\) which satisfies
\(m(R) = \langle n(R) \rangle\). Examples:

\begin{itemize}
\item
  If
  \(m(R) = (u_2 - u_1) \times \int_{w_1}^{w_2} \frac{\beta w^{-1-\beta}}{\Gamma(1-\beta)}\, dw\),
  then \(Z_u\) has independent and identically distributed increments,
  i.e.~it is a Levy flight.
\item
  Letting
  \(m(R) = (u_2 - u_1) \times \int_{w_1}^{w_2} \frac{\beta w^{-1-\beta} e^{-\theta w}}{\Gamma(1-\beta)}\, dw\)
  results in \(Z_u\) being a \emph{tempered stable Levy flight} with
  tempering parameter \(\theta \ge 0\).
\end{itemize}

A dependence of the Levy measure on the position \(Y_u\) of the walker
can be achieved via letting \[
m(R) = \int_{u_1}^{u_2} \int_{w_1}^{w_2} \nu(w |Y_u)\,dw\, du
\]

for some \emph{Levy measure} with density \(\nu(w | y)\), which may vary
with \(y\). Recall that a Levy measure is defined by the requirement \[
\int_0^\infty \min\{1, w\} \nu(w | y)\,dw < \infty.
\] For instance, letting the fractional exponent \(\beta(y) \in (0,1]\)
depend on space, choosing
\(\nu(w | y) = \frac{\beta(y) w^{-1-\beta(y)}}{\Gamma(1-\beta(y))}\)
results in \(Z_u\) having independent increments, which follow the
stable distribution with continuously varying exponent \(\beta(Y_u)\)
(Straka 2018).

It will be convenient to introduce the space-dependent tail function of
the Levy measure \[
\overline \nu(w | y) := \int_w^\infty \nu(w | y)\,dw, \quad w > 0.
\] and its Laplace transform \[
\hat{\overline \nu}(\lambda | y) = \int_0^\infty \overline \nu(w|y)\,e^{-\lambda w}\,dw.
\] We can then define the \emph{renewal function} \(V(y,s)\) via its
Laplace transform \[
\hat V(y,\lambda) := \int_0^\infty V(y,s)\,e^{-\lambda s}\,ds
= \frac{1}{\lambda [d(y) + \hat{\overline \nu}(\lambda | y)]}
\] The renewal function represents the mean occupation time of \(Z_u\)
conditional on \(y\) (Meerschaert and Straka 2012); that is, \(V(y,s)\)
is the mean amount of auxiliary time (\(u\)) for which \(Z_u < s\), if
\(y\) is frozen.

As shown by Baeumer and Straka (2016), the Fokker--Planck equation with
memory \eqref{eq:FFPE} has, under certain continuity conditions on the
four coefficient functions, a unique solution \(P(y,t)\). This solution
coincides with the probability distribution at time \(t\) of the
subordinated process

\begin{align}
\label{eq:CTRWlimit}
X(t) := Y_{E(t)}, \quad E(t) := \inf\{u: Z_u > t\}.
\end{align}

\(X(t)\) is also called a CTRW limit or the \emph{continuum limit} of
the CTRW.

\subsection*{Coefficient
representation}\label{coefficient-representation}
\addcontentsline{toc}{subsection}{Coefficient representation}

We note that the 4-tuple

\begin{align} \label{eq:coeff-tuple}
\left(a(x,t), b(x,t), d(x), \overline \nu(w, x)\right)
\end{align}

concisely represents the Langevin process \eqref{eq:dY}--\eqref{eq:dZ}.
However, the representation is only unique up to a multiplicative
factor: if every element in \eqref{eq:coeff-tuple}, say, doubled, then
the speed of \((Y_u, Z_u)\) is doubled. But, this has no effect on the
distribution of the points that are traversed by \((Y_u,Z_u)\), and
hence does not affect the distribution of the trajectories \(X(t)\).
This would remain true even if the speed varied with the location
\((x,t)\) of \((Y_u,Z_u)\).

Assuming that the coefficients are all bounded functions in \((x,t)\),
we hence divide by a large enough number so that \(a(x,t) < 1\) for all
\((x,t)\). (At \(a(x,t) = 1\), numerical instabilities may occur, which
are smoothed out if e.g. \(a(x,t) < 0.9\) throughout the domain.) In the
derivation of our algorithm, we will transform the tuple
\eqref{eq:coeff-tuple} as follows: Define \(\theta(x) \in [0, 1)\) via
\(d(x) = \theta(x) / (1-\theta(x))\). Then multiply the tuple
\eqref{eq:coeff-tuple} by \((1-\theta(x))\), to get the transformed
tuple

\begin{align} \label{eq:coeff-tuple-trans}
\left((1-\theta(x)) a(x,t), (1-\theta(x)) b(x,t), \theta(x), 
(1-\theta(x)) \overline \nu(w, x)\right).
\end{align}

Hence if we assume that \(d(y)\) is bounded, then we may also assume
WLOG that \(0 \le d(x) \le 1\) and \(a(x,t) < 1\).

Finally, we add the technical but non-restrictive condition

\begin{align} \label{eq:blowup}
\overline \nu(w|y) \le G(y) \, \frac{w^{-\beta(y)}}{\Gamma(1-\beta(y))}, 
\quad w \downarrow 0,
\end{align}

for some bounded function \(G(y)\), which prevents the Levy measure from
blowing up in regions where \(\beta(y) \uparrow 1\), see Lemma
\ref{lem:psi}.

\subsubsection*{Remark}\label{remark}
\addcontentsline{toc}{subsubsection}{Remark}

So-called Lipschitz and Growth conditions on the coefficients
\(\left(a(x,t), b(x,t), d(x), \overline \nu(w, x)\right)\) guarantee the
existence of the Langevin process \((Y_u, Z_u)\) (Applebaum 2009,
Chapter 6). These entail continuity of the parameters. It is generally
difficult to ensure existence of \((Y_u, Z_u)\) without these
conditions. For a recent approach of constructing the CTRW limit
\(X(t)\) without this condition, see Orsingher, Ricciuti, and Toaldo
(2018).

\section{Discrete Langevin Dynamics}\label{discrete-langevin-dynamics}

Let \(c > 0\) be a scaling parameter, and define a spatio-temporal grid
\(\#\) with spacings \(\chi \sim c^{-1/2}\) and \(\tau = 1/c\). Assuming
for simplicity that space is one-dimensional, the grid is embedded in
space-time \(\mathbb R \times [0, \infty)\). In this section we define
for each \(c > 0\) a Langevin process \((Y^{(c)}_u, Z^{(c)}_u)\) with
state space \(\#\) such that as \(c \to \infty\),
\((Y^{(c)}_u, Z^{(c)}_u)\) converges to \((Y_u, Z_u)\) in the sense of
stochastic processes.

It is clear that \((Y^{(c)}_u, Z^{(c)}_u)\) must be a jump process
hopping on \(\#\). Since \(Y_u\) has continuous sample paths, nothing is
gained by allowing \(Y^{(c)}_u\) to jump to non-neighbouring lattice
sites. Also, since \(Z_u\) is increasing, \(Z^{(c)}_u\) need not jump
backwards. It is helpful to view the sequence of grid points traversed
by \((Y^{(c)}_u, Z^{(c)}_u)\) as locations and times of a walker
performing a DTRW (discrete time random walk), with jumps and waiting
times given by the increments of \(Y^{(c)}_u\) resp. \(Z^{(c)}_u\).

\subsection{Waiting time distribution}\label{waiting-time-distribution}

We define the discrete waiting time distribution
\(\psi^{(c)}(j\tau | x)\) as a mixture of a ``local'' and a ``nonlocal''
component:

\begin{align}
\psi^{(c)}(j\tau | x) := \theta(x) \psi^{{(c)}}_{\rm loc}(j\tau | x) 
+ (1-\theta(x)) \psi^{(c)}_{\rm nonloc}(j\tau|x), \quad j=1,2,\ldots
\label{eq:def-psi}
\end{align}

where, by definition, \(0 \le \theta(x) \le 1\). The local part is
simply deterministic, with all mass at \(\tau\), that is
\(\psi^{(c)}_{\rm loc}(\tau | x) = 1\) and
\(\psi^{(c)}_{\rm loc}(k\tau | x) = 0\) for \(k=2, 3, \ldots\). The
nonlocal part is the truncated, normalized and discretized Lévy measure:
First, define the function \[
H^{(c)}(w | x) = 1 \wedge \frac{\overline \nu(w|x)}{c},
\] where \(a \wedge b := \min\{a, b\}\). For convenience, we say that
\(\overline \nu(w|x) = \infty\) if \(w \le 0\). Then, define

\begin{align}
w_\tau := j\tau, \quad \text{ where }  \quad j\tau \le w < (j+1)\tau.
\label{eq:w-tau}
\end{align}

Finally, note that
\(\Psi^{(c)}_{\rm nonloc}(w|x) := H^{(c)}(w_\tau | x)\) is piecewise
constant with jumps in \(\tau, 2\tau, \ldots\), and decreasing from
\(1\) to \(0\). We take this function to be the tail function of
\(\psi^{(c)}_{\rm nonloc}(w|x)\), that is,

\begin{align}
\psi^{(c)}_{\rm nonloc}(j\tau | x) = H^{(c)}((j-1)\tau | x) - H^{(c)}(j\tau | x), 
k = 1, 2, \ldots.
\end{align}

We then have \(\psi^{(c)}(0\tau | y) = 0\), meaning that waiting times
are always strictly positive.

\subsection{Jump distribution}\label{jump-distribution}

We assume that the DTRW jumps can have one of the three values
\(\{-\chi, 0, +\chi\}\), where \(\bar a = \sup\{a(x,t)\}\) and
\(\chi = (\bar a / c)^{1/2}\). The probabilities to jump left, to
``self-jump'' (i.e.~jump back to the original location), and to jump
right, are given by

\begin{gather*}
\ell^{(c)}(x,t) = \frac{a(x,t) - \chi b(x,t)}{2 \bar a},
\quad
n(x,t) = 1 - a(x,t)/\bar a,
\quad
r^{(c)}(x,t) = \frac{a(x,t) + \chi b(x,t)}{2 \bar a}
\end{gather*}

where \(x\) is the location of the walker before the jump, and \(t\) is
the time at which the jump occurs. In order for \(r, n\) and \(\ell\) to
be between \(0\) and \(1\), we need \(\chi\) to be small enough so that
\[
\chi |b(x,t)|  \le a(x,t), \quad (x,t) \in \mathbb R \times [0,\infty).
\]

\subsection{Convergence}\label{convergence}

At scale \(c\), the probabilies \(\psi^{(c)}(j\tau | y)\) and
\(\ell^{(c)}(x,t)\), \(n(x,t)\) and \(r^{(c)}(x,t)\) define a jump
kernel on \(\#\), which defines the distribution of jump \(z\) and
waiting time \(w\) given the current location of the walker at \(x\) at
time \(t\):

\begin{align}
\begin{split}
& K^{(c)}(z,w|x,t)
\\
&= \left[r^{(c)}(x,t+w)\delta_{+\chi}(z)
+ n(x, t+w) \delta_0(z)
+ \ell^{(c)}(x, t+w) \delta_{-\chi}(z)\right]
\psi^{(c)}(w|x).
\label{eq:K}
\end{split}
\end{align}

Note that we evaluate the jump probabilities at the end \(t+w\) of a
waiting time, as is common for CTRWs. Th. 2.1 in (Straka 2018) specifies
conditions on \(K^{(c)}(z,w|x,t)\) which imply the convergence of
\((Y^{(c)}_u, Z^{(c)}_u)\) to \((Y_u, Z_u)\) and which we repeat here
for convenience:

\begin{align} \label{eq:cond1}
\lim_{\epsilon \downarrow 0} \lim_{c \to \infty}
\iint\limits_{|z|< \epsilon,\,0<  w < \epsilon} z cK^{(c)}(z,w | x,s)\,dz\,dw &= b(x,s)
\\ \label{eq:cond2}
\lim_{\epsilon \downarrow 0} \lim_{c \to \infty}
\iint\limits_{|z|< \epsilon, \,0<w < \epsilon} z^2 cK^{(c)}(z,w | x,s)\,dz\,dw &= a(x,s)
\\ \label{eq:cond3}
\lim_{\epsilon \downarrow 0} \lim_{c \to \infty}
\iint\limits_{|z|< \epsilon, \,0<w < \epsilon} w cK^{(c)}(z,w | x,s)\,dz\,dw &= \theta(x)
\\
\label{eq:cond4}
\lim_{c \to \infty}
\iint\limits_{z \in \mathbb R, w \ge 0} g(z,w) cK^{(c)} (z,w | x,s)\,dz\,dw &= \int_{w > 0} g(0,w) \nu(w|x)\,dw
\end{align}

for any bounded continuous function \(g(z,w)\) which vanishes in a
neighbourhood of the origin. We give calculations in the appendix which
confirm that the above four conditions indeed hold for
\(K^{(c)}(z,w|x,t)\) as defined in \eqref{eq:K}.

\subsubsection*{Remark}\label{remark-1}
\addcontentsline{toc}{subsubsection}{Remark}

The alternative kernel

\begin{align} \label{eq:Kalt}
  K^{(c)}(z,w|x,t) = \left[r^{(c)}(x,t)\delta_{+\chi}(z)
  + n(x, t) \delta_0(z)
  + \ell^{(c)}(x, t) \delta_{-\chi}(z)\right]
  \psi^{(c)}(w|x).
\end{align}

also satisfies \eqref{eq:cond1} -- \eqref{eq:cond4}. The difference to
\eqref{eq:K} is that the probabilities \(r^{(c)}(x,t), \ell^{(c)}(x,t)\)
and \(n(x,t)\) are evaluated at the \emph{beginning} of a waiting time,
rather than the end. As investigated by C. N. Angstmann, Donnelly,
Henry, Langlands, et al. (2015), this difference vanishes in the limit
as \(c \to \infty\).

\section{Semi-Markov numeric scheme}\label{semi-markov-numeric-scheme}

As described at the beginning of Section 3, the discrete Langevin
process \((Y^{(c)}_u, Z^{(c)}_u)\) has an embedded DTRW, for which we
write \(X^{(c)}(t)\). By Theorem 2.2 in Straka (2018),

\begin{equation}
X^{(c)}(t) \text{ converges to the CTRW continuum limit process } X(t)
\label{eq:CTRW-J1}
\end{equation}

from \eqref{eq:CTRWlimit}. (Convergence here means weak convergence with
respect to the \(J_1\) topology of right-continuous sample paths with
left-hand limits, see Whitt (2001).) For large \(c\), the probability
distributions of \(X^{(c)}(t)\) may hence be taken as approximations of
\(P(y,t)\). In this section, we derive master equations for the
probability distributions of \(X^{(c)}(t)\).

\subsection{Semi-Markov property}\label{semi-markov-property}

A DTRW starting at \(x\) at time \(t\) is defined by the jump kernel
\eqref{eq:K} as follows: first, a waiting time is drawn from the
distribution \(\psi^{(c)}(w|x)\); then a jump left or right or a
self-jump is drawn from the probabilities \(\ell^{(c)}(x,t+w)\),
\(r^{(c)}(x,t+w)\) and \(n(x,t+w)\). The Semi-Markov approach embeds
\(X^{(c)}(t)\) into a Markov process (Meerschaert and Straka 2014):
Define the \emph{age} of a walker as the time that has passed since he
last arrived at his current location. In each timestep \(\tau\), either
the waiting time has not expired yet, in which case no jump occurs and
age is increased by \(\tau\); or age is reset to \(0\) and a jump
occurs. Since this recipe determines the future evolution of position
and age based on only the current position and age, the process is
Markovian, and it is straightforward to derive master equations.

Recall that a waiting time \(W\) at a spatial lattice point \(i\chi\) is
drawn from \(\psi^{(c)}(w | i\chi)\) and thus satisfies \[
\mathbf P(W > j\tau) = H^{(c)}(j\tau | i\chi) =: h_{i,j}.
\] Conditional on \(W > j\tau\), the probability that
\(W > (j + 1) \tau\) is \[
\mathbf P(W > (j + 1)\tau | W > j \tau) = h_{i,j+1} / h_{i,j}.
\] That is, if at time \(k\tau\), position and age are
\((x_k, v_k) = (i,j)\), then at time \((k+1)\tau\) the pair
\((x_{k+1}, v_{k+1})\) is equal to

\begin{itemize}
\tightlist
\item
  \((x_k, v_k + 1)\) with probability \(h_{i,j+1} / h_{i,j}\), and
\item
  \((x_k + \zeta, 0)\) with probability \(1 - h_{i,j+1} / h_{i,j}\),
\end{itemize}

where \(\zeta \in \{-1, 0, +1\}\) with probabilities
\(\ell^{(c)}(i\chi, (k+1)\tau)\), \(n(i\chi, (k+1)\tau)\) and
\(r^{(c)}(i\chi, (k+1)\tau)\).

The above dynamics uniquely determine the stepwise evolution of
\((x_k, v_k)\). We write \(\xi^k_{i,j} = \mathbf P(x_k = i, v_k = j)\)
for the probability distribution of \((i,j)\) at time \(k\). The master
equations for \(\xi^k_{i,j}\) then read:

\begin{align}
\label{eq:GME1}
\xi^{k+1}_{i,j} &= \frac{h_{i,j}}{h_{i,j-1}}\, \xi^k_{i,j-1}, \quad 1 \le j < J-1,
\\
\label{eq:GME2}
\xi^{k+1}_{i,0} &= \sum_{j=0}^J\left(1 - \frac{h_{i,j+1}}{h_{i,j}}\right)
(\ell^k_{i+1} \xi^k_{i+1, j} + r^k_{i-1} \xi^k_{i-1,j}
  + n^k_{i,j} \xi^k_{i,j})
\end{align}

The line \eqref{eq:GME1} states that for a walker to have age
\(j \ge 1\), it must have had age \(j - 1\) in the previous time step,
and not jumped. The line \eqref{eq:GME2} states that for a walker to
have age \(j = 0\), it must have jumped to its location \(i\) in the
previous time step, from a neighbouring lattice site or from \(i\)
itself. The probability mass of all walkers jumping from site \(i\)
during time step \(k \to k+1\) is
\(\sum_{j=0}^J \left(1 - h_{i,j+1} / h_{i,j}\right) \xi^k_{i,j}\), which
is redistributed according to the probabilities \(r^{k+1}_{i,j}\),
\(\ell^{k+1}_{i,j}\) and \(c^{k+1}_{i,j}\). This interpretation shows
that \eqref{eq:GME1}--\eqref{eq:GME2} \textbf{conserve probability
mass}.

Iterating the equation pair \eqref{eq:GME1}--\eqref{eq:GME2} from some
initial condition computes the evolution of the joint probability
distribution of position and age. The marginal distribution of the
position is calculated simply via \[
\mathbf P(X^{(c)}_t = i\chi) =: \rho^k_i = \sum_{j=0}^J \xi^k_{i,j},
\quad k = \lfloor t /\tau \rfloor.
\] Here we note that
\(X^{(c)}(t) = X^{(c)}(t_\tau) = X^{(c)}(\lfloor t/\tau \rfloor \tau)\),
where \(t_\tau\) is the left-nearest lattice point defined exactly as
\(w_\tau\) in \eqref{eq:w-tau}.

\subsection{Boundary conditions}\label{boundary-conditions}

In practice, one can only allocate a finite number \(J\) of points to
the lattice of ages. If we cannot allocate \(\lfloor T/\tau \rfloor\)
lattice points, where \(T\) is the largest time of interest, then it is
possible that the age of walkers may reach the end of the lattice. In
this case, and if the walker does not jump in the next time step, we do
not increase its age any further, until it eventually does jump: \[
\xi^{k+1}_{i,J} = \frac{h_{i,J}}{h_{i,J-1}} \xi^k_{i,J-1}
+ \frac{h_{i,J+1}}{h_{i,J}} \xi^k_{i,J},
\] The first summand being walkers whose age has reached \(J\) in the
current time step, and the second summand being walkers of age \(J\) who
do not jump in the current time step. Finally, assuming that the spatial
coordinates of the lattice go from \(-I\) to \(I\), we implement Neumann
boundary conditions by placing a walker back on the boundary whenever it
would otherwise have jumped off the lattice, that is:

\begin{align}
  \ell^k_{-I} &= 0, & n^k_{-I} &= \ell(-I\chi, k\tau) + n(-I\chi, k\tau),
  & r^k_{-I} &= r(-I\chi, k\tau),
  \\
  \ell^k_I &= \ell(I \chi, k\tau), & n^k_{I} &= n(I\chi, k\tau) + r(I\chi, k\tau),
  & r^k_{I} &= 0
\end{align}

\subsection{Properties of the
algorithm}\label{properties-of-the-algorithm}

\subsubsection*{Positivity}\label{positivity}
\addcontentsline{toc}{subsubsection}{Positivity}

From \eqref{eq:GME1}--\eqref{eq:GME2}, it is evident that the
\(\xi^k_{i,j}\) are necessarily non-negative, and hence the solution
\(\rho^k_i\) cannot be negative.

\subsubsection*{Consistency of the
algorithm}\label{consistency-of-the-algorithm}
\addcontentsline{toc}{subsubsection}{Consistency of the algorithm}

Due to the convergence \eqref{eq:CTRW-J1}, we have

\begin{align} \label{eq:consistency}
  \sum_{i=-I}^I f(i\chi) \rho^{\lfloor t/\tau \rfloor}_i
  = \langle f(X^{(c)}_t) \rangle
  \longrightarrow \langle f(X_t) \rangle
  \quad \text{ as } c \to \infty,
\end{align}

for all bounded continuous real-valued \(f\) defined on \(\mathbb R\).
If the distribution of \(X_t\) has a probability density, then the above
convergence also holds if \(f\) is an indicator function of an interval
\((a,b)\), and reads

\begin{align}
  \sum_{a < i\chi < b} \rho_i^{\lfloor t / \tau \rfloor}
  \longrightarrow
  \mathbf P(a < X_t < b) \quad \text{ as } c \to \infty.
\end{align}

\subsubsection*{Equivalence with DTRW
approach}\label{equivalence-with-dtrw-approach}
\addcontentsline{toc}{subsubsection}{Equivalence with DTRW approach}

The Discrete Time Random Walk algorithm by C. N. Angstmann, Donnelly,
Henry, and Nichols (2015) assumes discrete waiting times with the Sibuya
distribution, whose tail function \(\Psi(n)\) has the asymptotics
\(\Psi(n) \sim n^{-\beta}\). In \eqref{eq:GME2}, see that we have
\(\xi^k_{i,j} = \xi^{k-j}_{i,0} h_{i,j}\), by telescoping
\eqref{eq:GME1} and \(h_{i,0} = 1\). Hence \eqref{eq:GME2} rewrites to
\[
\xi^{k+1}_{i,0} = \sum_{j=0}^J (h_{i,j} - h_{i,j+1})
(\ell^{k+1}_{i+1} \xi^{k-j}_{i+1, 0} + r^{k+1}_{i-1} \xi^{k-j}_{i-1,0} + c^{k+1}_{i,j} \xi^{k-j}_{i,0}),
\] assuming that \(h_{i,j}\) is constant in \(i\) (homogeneous waiting
times). Since \(h_{i,j} - h_{i,j+1}\) is the probability of a waiting
time being \(j+1\), one sees the equivalence of methods by comparing
with Equation (16) in C. N. Angstmann, Donnelly, Henry, and Nichols
(2015), if we choose \(h_{i,j} = \Psi(j)\).

\section{Examples}\label{sec:examples}

Within our unifying semi-Markov framework, we may approximate
probability distributions of a great variety of CTRW limits. We study
several examples.

\subsection{Continuous interface
problem}\label{continuous-interface-problem}

\begin{figure}
\centering
\includegraphics{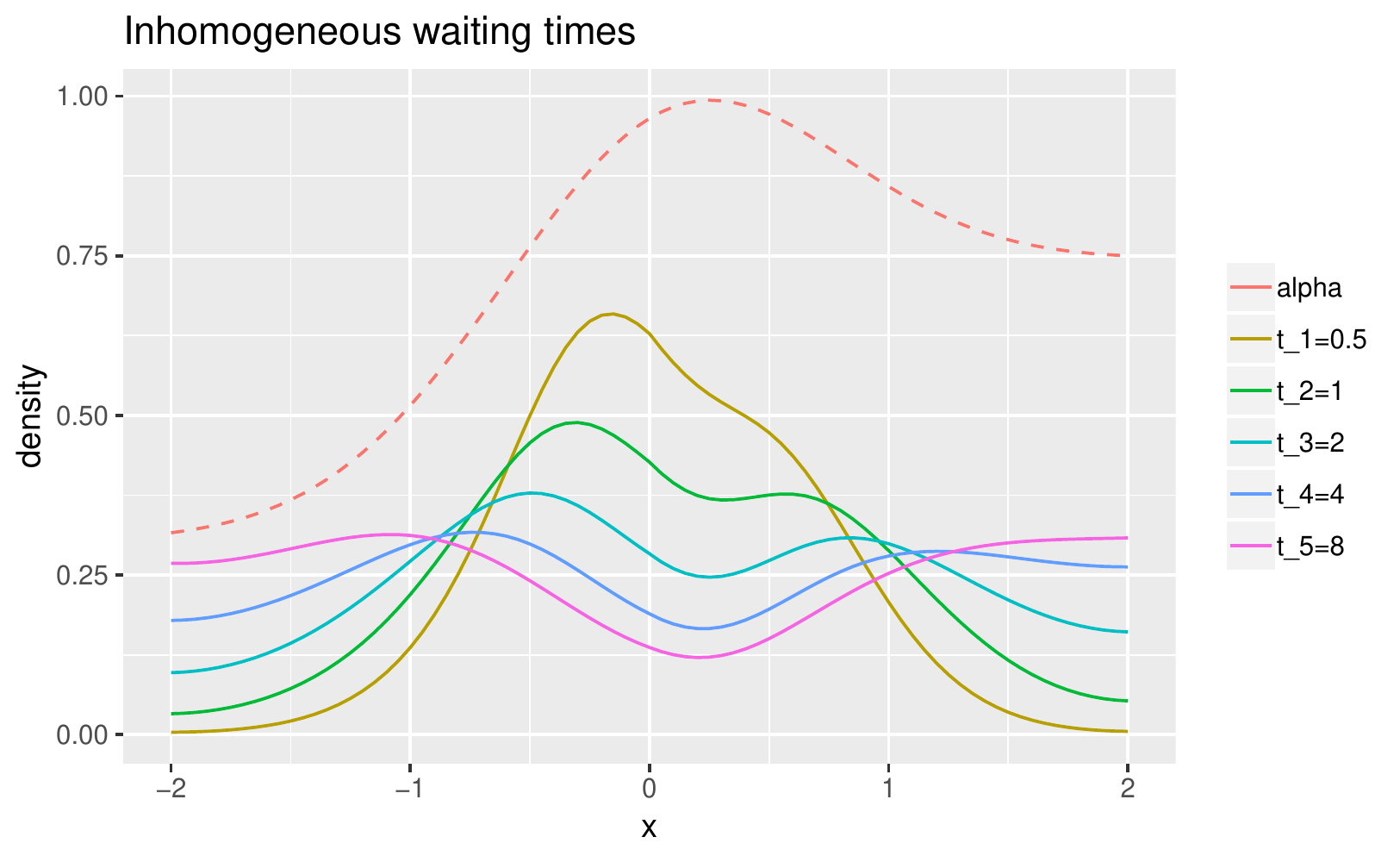}
\caption{\label{fig:interface}Continuous interface problem. Coefficients
are as given in the text, and \(c = 400\).}
\end{figure}

Korabel and Barkai (2010) have studied a one-dimensional subdiffusive
lattice with exponent \(\beta = 0.3\) for \(x<0\) and \(\beta = 0.75\)
for \(x>0\), where at the interface (\(x=0\)) the waiting time is
exponentially distributed. Even if particles are biased to jump to the
right at \(x=0\) and thus the net drift becomes positive, in the
long-time limit \emph{all particles end up in the left half}.

Here we consider a continuous medium that mimics this setup with the
coefficients \((\overline \nu(w|x), d(x), a(x,t), b(x))\) chosen as
follows:

\begin{align*}
\overline \nu(w|x) &= \frac{w^{-\alpha(x)}}{\Gamma(1-\alpha(x))} \text{ where }
\alpha (x) = 0.45e^{-x^2} + 0.3 + 0.45/(1+e^{-2x}), 
\\
d(x) &\equiv 0, 
\quad 
a(x,t) \equiv 1, 
\quad 
b(x,t) = 0.1 * \phi(x | 0, 0.2)
\end{align*}

where \(\phi(x | \mu, \sigma)\) denotes the probability density of the
Gaussian distribution with mean \(\mu\) and standard deviation
\(\sigma\). Note that \(\alpha(x,t)\) is chosen so that it approaches
\(0.3\) for large negative \(x\), \(0.75\) for large positive x and
remains just under \(1\) near \(x=0\).

Figure \ref{fig:interface} shows the evolution of the density \(P(y,t)\)
with a delta function initial condition. At small times we observe two
peaks reflecting the trapping that occurs either side of the interface.
For late times, one begins to see the aggregation of all particles
towards the left hand side (\(x<0\)) where trapping is stronger (Savov
and Toaldo 2018; Fedotov and Falconer 2012).

\begin{figure}
\centering
\includegraphics{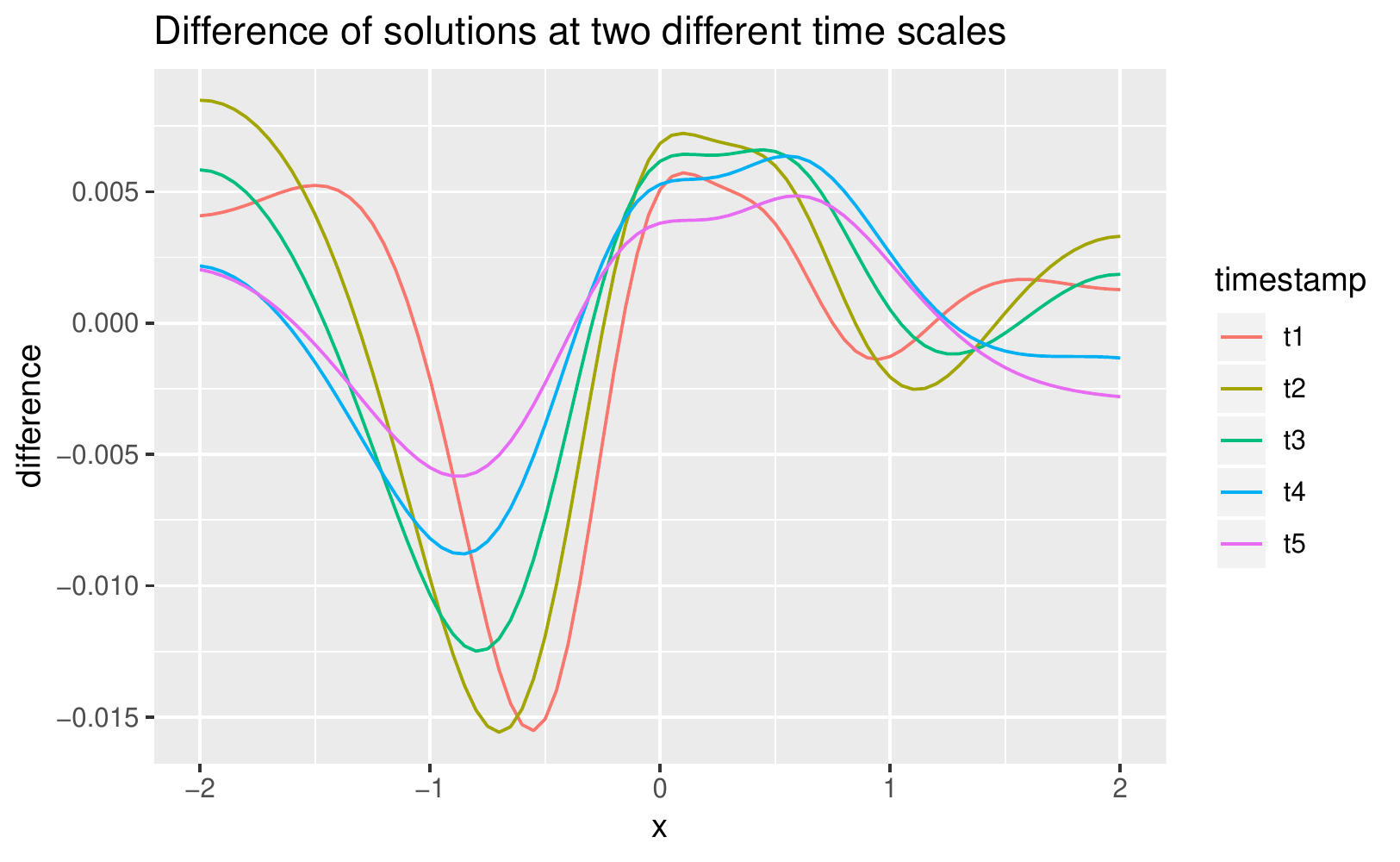}
\caption{\label{fig:difference}Absolute difference between densities,
calculated at corresponding timestamps, for two choices of time scale
\(T_0 = 1\) and \(T_0 = 2\).}
\end{figure}

Straka (2018) shows that changing time units from \(T_0 = 1\) to
\(T_0 = 2\) results in the the updated diffusivity and drift
coefficients

\begin{align*}
a_{\beta(x)}(x,t) = \frac{a(x,t)}{T_0^{-\alpha(x)}}, 
\quad 
b_{\beta(x)}(x,t) = \frac{b(x,t)}{T_0^{-\alpha(x)}},
\end{align*}

leading to spatially inhomogeneous temporal scaling. We confirm this by
computing probability densities for the parameter tuple
\(\left( a_{\alpha(x)}(x,t), b_{\alpha(x)}(x), d(x), \overline \nu(w|x)\right)\),
at the timestamps multiplied by \(T_0 = 2\), and plotting the absolute
differences (Figure \ref{fig:difference}). The absolute values of
differences are mostly all below 0.015, and remain stable after 8 units
of time, indicating that indeed the same densities are calculated in
both cases.

\subsection{\texorpdfstring{Temporal drift
\(d(x)\)}{Temporal drift d(x)}}\label{temporal-drift-dx}

\begin{figure}
\centering
\includegraphics{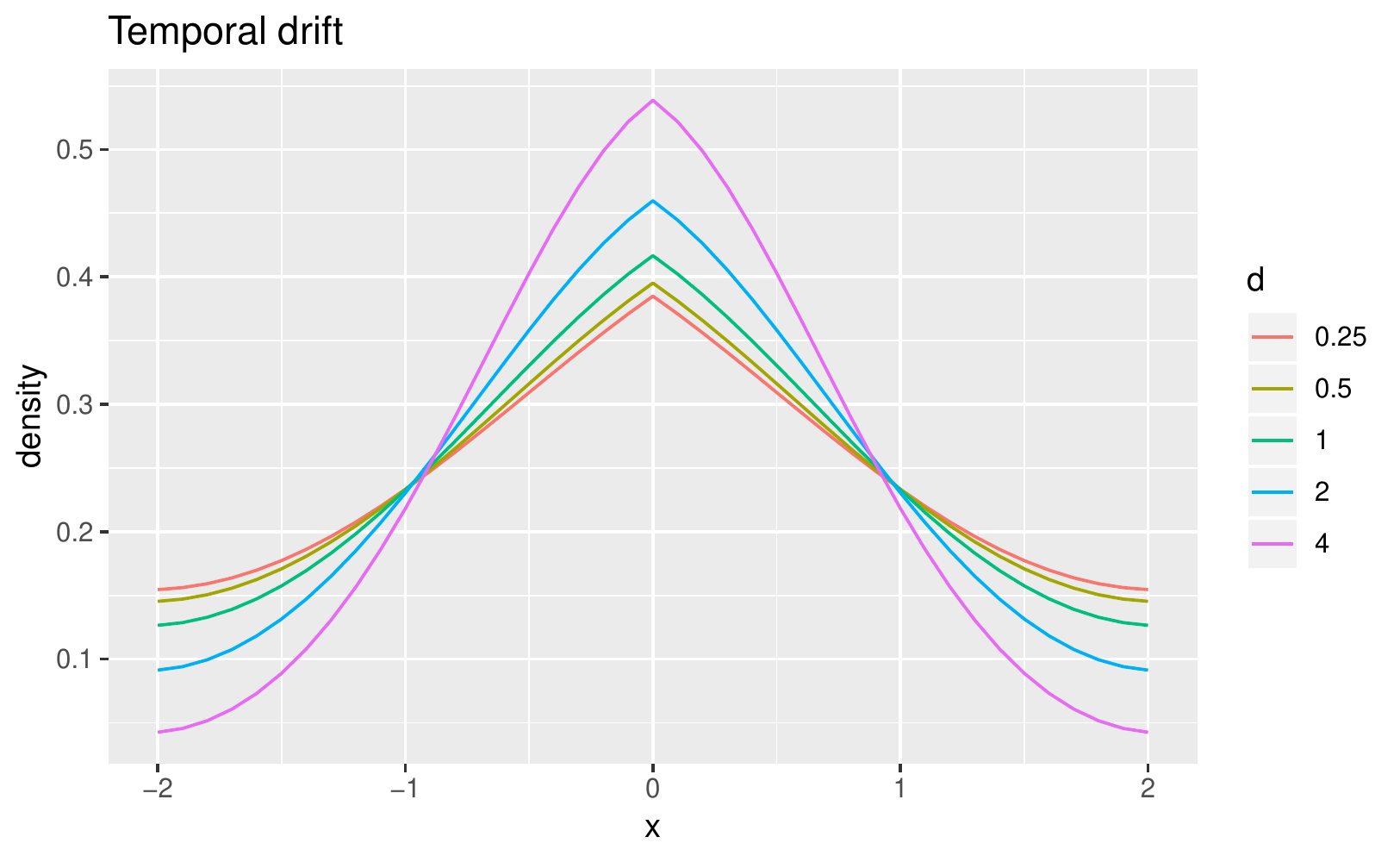}
\caption{\label{fig:temporal-drift}Increasing temporal drift \(d(x)\)
decreases the speed of the diffusion and increases resemblence to a
standard Gaussian process. Parameters:
\(a(x,t) = 0.8, b(x,t) = 0, \overline\nu(w|x)=w^{-0.7}/\Gamma(1-0.7), c=100, \tau = 1/100, \chi = 1/10\).}
\end{figure}

\begin{figure}
\centering
\includegraphics{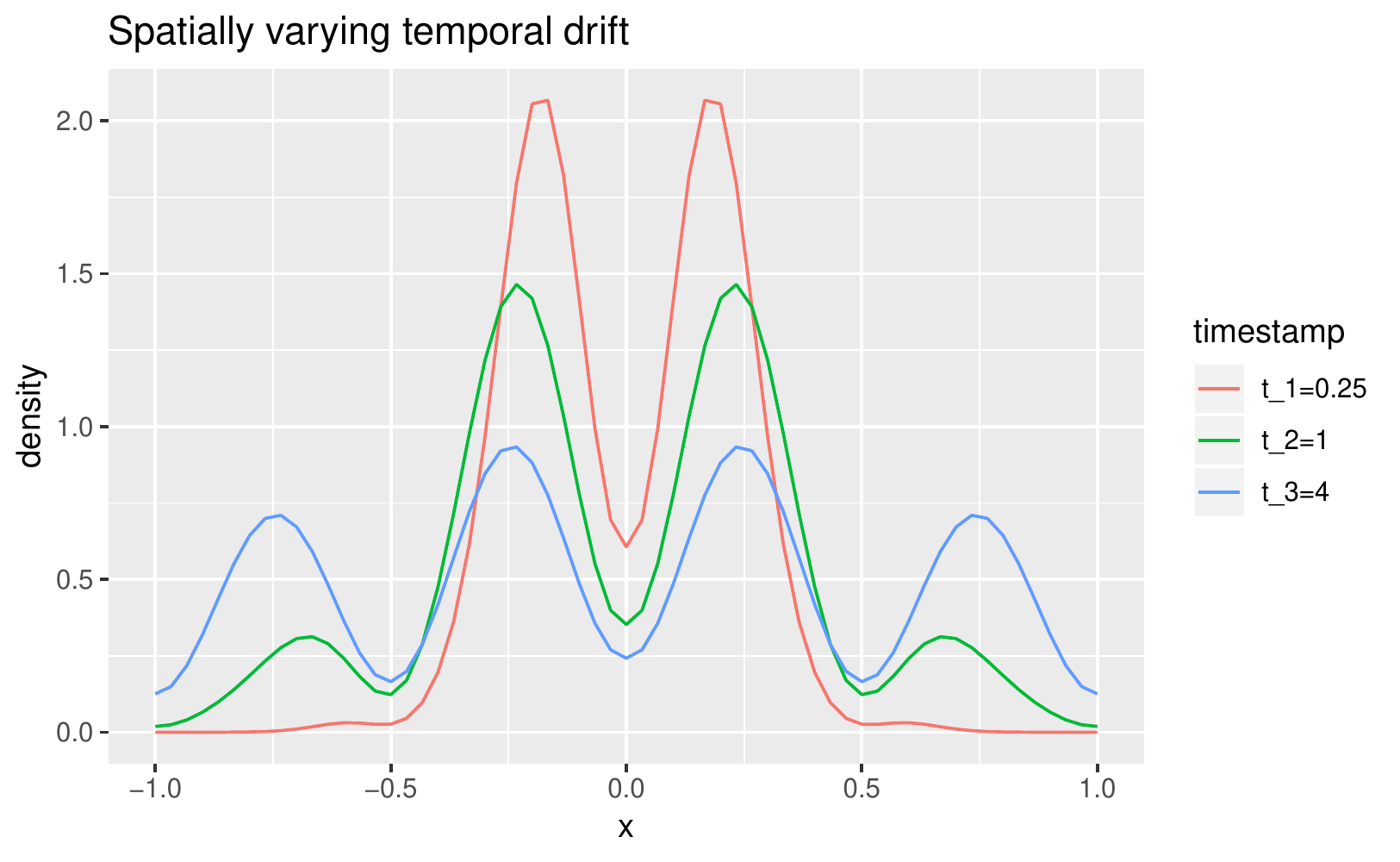}
\caption{\label{fig:varying-temporal-drift}A system with a spatially
varying temporal drift \(d(x) = 10 * \sin^2(2\pi x)\). Particles
accumulate in the slow patches where \(d(x)\) is high, while trapping is
homogeneous in space. Other parameters: \(a(x,t) = 0.9\),
\(b(x,t) = 0\), \(\overline\nu(w|x) = w^{-0.7}/\Gamma(1-0.7)\),
\(c = 900\).}
\end{figure}

CTRW limits with positive temporal drift \(d(x)\) as per representation
\eqref{eq:dY}--\eqref{eq:dZ} have been studied by Straka (2011): In the
case where \(Z_u\) is a \(\beta\) stable Lévy flight, \(Z_u\) grows
superlinearly at the rate \(u^{1/\beta}\) both in the short time limit
\(t \downarrow 0\) and the long time limit \(t \uparrow \infty\).
Accordingly, the inverse stable subordinator \(E(t)\) in
\eqref{eq:CTRWlimit} grows as \(\propto t^\beta\), also both in the
short time and long time limit. Adding a drift to \(Z_u\), e.g.
\(d(x) \equiv d > 0\), means that \(Z_u\) now grows linearly
\(\propto d\,u\) at short times. Accordingly, its inverse \(E(t)\) also
grows linearly as \(\propto t/d\) at short times. The growth behaviour
at late times of \(Z_u\) and \(E(t)\) remains dominated by large jumps
resp. long rests, and remains \(\propto u^{1/\beta}\) resp.
\(\propto t^\beta\). Hence the addition of the drift \(d > 0\) means
that the slope of \(E(t)\) is no longer infinite, and thus the speed of
\(E(t)\) is tempered at very short times. Figure
\ref{fig:temporal-drift} illustrates the effect of increasing the
temporal drift. As can be seen, the jump component of \(Z_u\) becomes
less pronounced as the temporal drift increases, increasing resemblance
to a Gaussian process and slowing down the dynamics. Figure
\ref{fig:varying-temporal-drift} shows anomalous diffusion with exponent
\(0.7\) with spatially varying temporal drift \(d(x)\). Particles
accumulate in patches of low mobility, corresponding to high \(d(x)\).

\subsection{Variably distributed fractional
order}\label{variably-distributed-fractional-order}

Anomalous diffusion with \emph{distributed order} assumes a mixing
probability distribution of the anomalous parameter \(\beta\) with
density \(p(\beta)\) on the interval \((0,1]\). As illustrated by Sandev
et al. (2015), the position of the distributed order fractional operator
is decisive for the long-term dynamics. The ``natural form'' uses the
Caputo fractional derivative: \[
\int_0^1 p(\beta) _CD_t^\beta P(y,t) \,d\beta = \frac{1}{2}\frac{\partial^2}{\partial y^2} P(y,t)
\] Here the mean squared displacement grows proportionally to
\(t^{\beta_1}\) for early times and proportionally to \(t^{\beta_2}\)
for late times, where \(\beta_1\) is at the left end of the support of
\(p(\beta)\) and \(\beta_2\) at the right end. The opposite behaviour
occurs for the ``modified form'', with Riemann-Liouville fractional
derivative: \[
P(y,t) = \int_0^1 p(\beta) _{RL}D_t^{1-\beta}\frac{1}{2}\frac{\partial^2}{\partial y^2} P(y,t)\,d\beta
\] The FFPE for CTRW limits \eqref{eq:FFPE} can be rewritten to the
natural form, assuming that all coefficients
\((a, b, d, \overline \nu(w)\) are constant (compare with Eq.(3.8) in
Straka 2018 with delta-function initial condition and
\(\overline \nu(t) = t^{-\beta}/\Gamma(1-\beta)\)):

\begin{align} \label{eq:natural-mixture}
d\, \frac{\partial}{\partial t} P(y,t) + _CD_t^\beta P(y,t)
= \mathcal L^* P(y,t)
\end{align}

which represents a mixture of the two orders \(1\) and \(\beta\), with
weights \(d/(d+1)\) and \(1/(d+1)\), after normalization.

\begin{figure}
\centering
\includegraphics{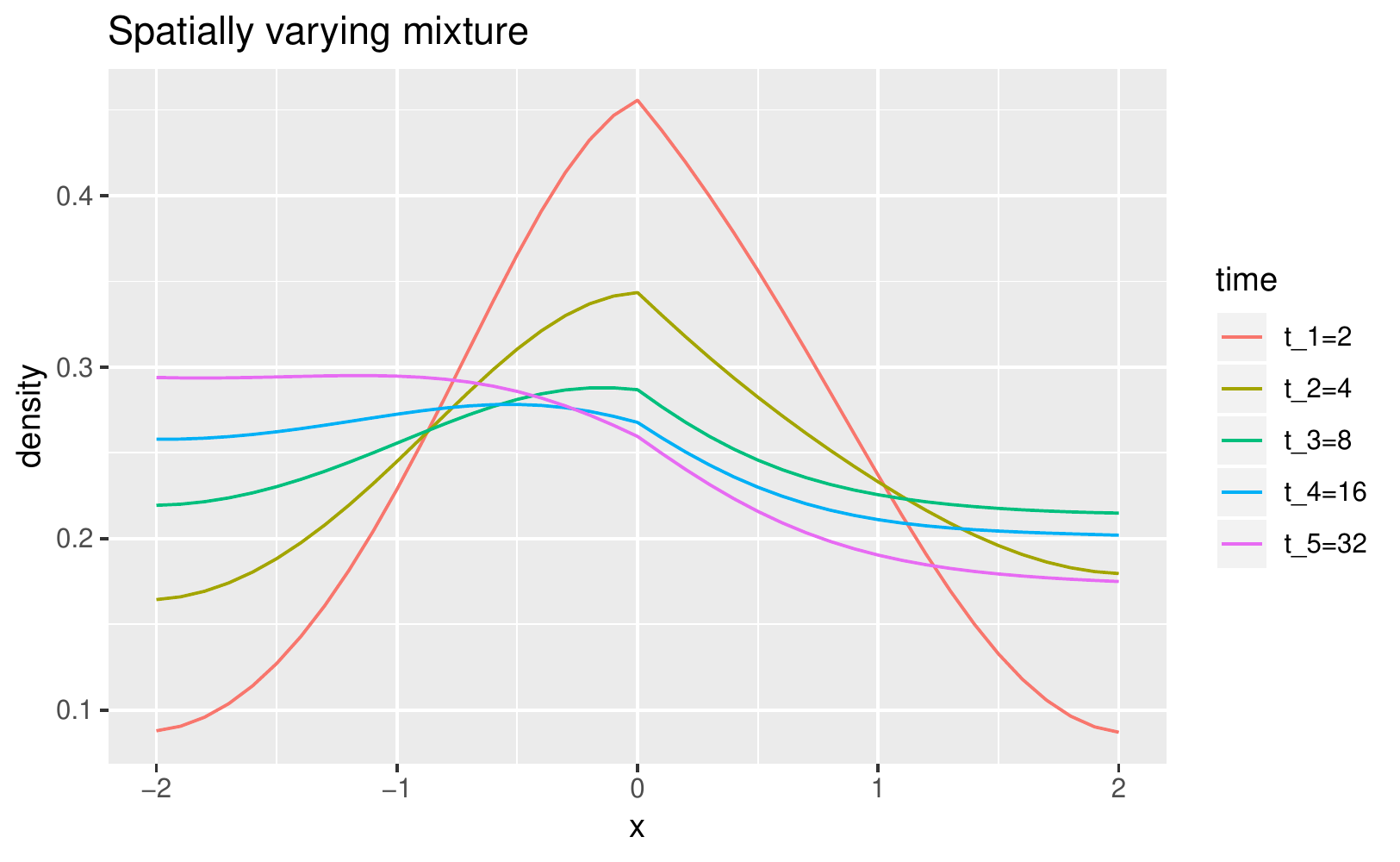}
\caption{\label{fig:varying-mixture}A variable mixture of subdiffusion
(\(\beta_1 = 0.7\)) and diffusion (\(\beta_2 = 1\)). The two weights add
to \(1\), and the weight for \(\beta_2\) equals the logistic function
with scale \(1/2\), increasing from \(0\) on the far left to \(1\) on
the far right.}
\end{figure}

We now vary the weights of the two orders in space: Assume a logistic
weight \(p(y) = 1/(1+\exp(-2y)\) with scale \(0.5\) for the exponent
\(1\), and the weight \(1-p(y)\) for the exponent \(\beta = 0.7\). Then
the dynamics are diffusive on the far right-hand side, subdiffusive on
the far left-hand side, and mixed at the interface near \(0\), with
continuous interpolation between the two regimes. This is summarized in
the coefficient tuple \[
\left( 0.9, 0, p(y), (1-p(y)) w^{-0.7} / \Gamma(1-0.7) \right).
\] Note however that the CTRW limit specified by this tuple is
\textbf{not} governed by \eqref{eq:natural-mixture} with weights of
\(1\) and \(\beta\) replaced by \(p(y)\) and \(1-p(y)\), since the
derivation of this equation assumes constant coefficients. We deem it
unlikely that a Caputo-type governing equation of the above dynamics
exists. The evolution of a system with point mass initial condition at
the interface \(y=0\) is illustrated in Figure
\ref{fig:varying-mixture}.

\subsection{Inverse subordinators}\label{inverse-subordinators}

\begin{figure}
\centering
\includegraphics{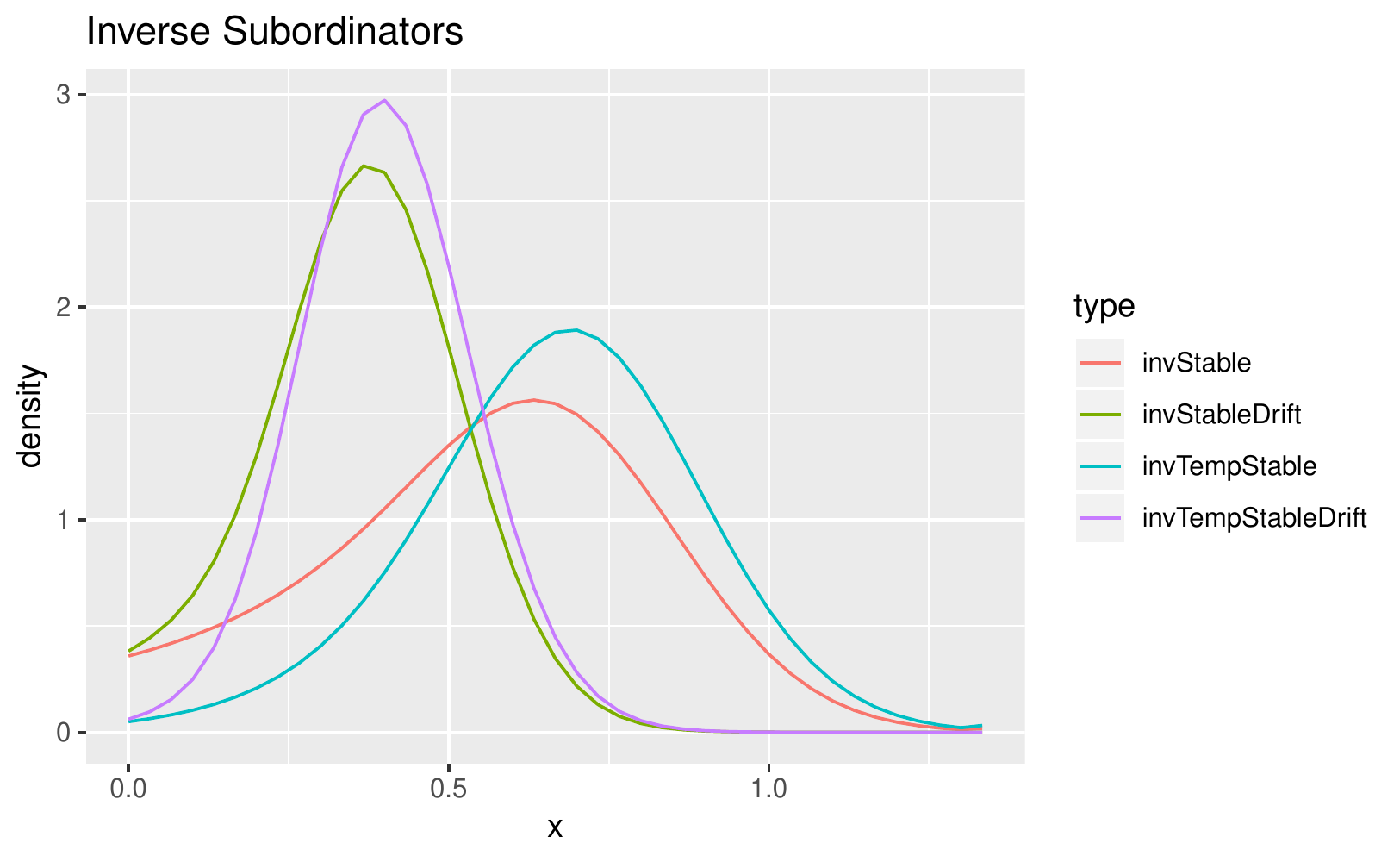}
\caption{\label{fig:invSubordinators}Probability densities for three
different types of inverse subordinators, evaluated at time \(t = 1\),
with \(c=900\).}
\end{figure}

The time-changing process \(E(t)\) from \eqref{eq:CTRWlimit} is
well-known in the statistical physics literature as an ``inverse
subordinator'', and denotes the random crossing time \(u\) of a level
\(t\) by the stable Levy flight \(Z_u\). Subordination is a widely used
method to simulate paths of CTRW limits, see e.g. (Meerschaert and
Straka 2013) for an overview. Alrawashdeh et al. (2017) study the
``inverse tempered stable subordinator'', i.e.~the level crossing time
for the \emph{tempered} stable Levy flight. This process is an important
tool for the study of ``tempered subdiffusion'', see e.g. Gajda and
Magdziarz (2010). Using our algorithm, we may compute probability
densities for \emph{any} inverse subordinator \(E(t)\), defined as the
level crossing time of any strictly increasing Levy flight.

To this purpose, observe that if \(Y_u = u\) is linear motion, then we
have \(X(t) = E(t)\) in \eqref{eq:CTRWlimit}. In order for \(Y_u = u\)
to hold, we simply let \(a(x,t) = 0\) and \(b(x,t) = 1\). In other
words, an inverse subordinator is a CTRW limit defined via a coefficient
tuple \[
\left( 0, 1, d, \overline \nu(w) \right)
\] A set of jump probabilities which achieves the limits in
\eqref{eq:cond1} and \eqref{eq:cond2} is \[
\ell^{(c)}(x,t) = 0, \quad n^{(c)}(x,t) = 1 - \chi, \quad r^{(c)} = \chi.
\] Figure \ref{fig:invSubordinators} shows probability densities of the
inverse stable, inverse tempered stable, inverse stable with drift and
inverse tempered stable with drift subordinators. These have coefficient
tuples with: \(a =0\), \(b = 1\); in the untempered resp. tempered case,
the tail of the Levy measure is \[
\overline \nu(w) = \frac{w^{-0.7}}{\Gamma(1-0.7)}
\quad \text{ resp.} \quad 
\overline \nu(w) = \frac{w^{-\beta} e^{-\gamma w} - \gamma \Gamma(1-\beta, w)}{\Gamma(1-\beta)}
\] where we set the tempering parameter \(\gamma = 1\); and for the
non-drift resp. drift case, \(d=0\) resp. \(d=1\). Since tempering makes
the Levy flight \(Z_u\) smaller, \(E(t)\) becomes larger. Moreover, a
drift \(d=1\) introduces the lower bound \(u \le Z_u\), which then
becomes an upper bound \(E(t) \le t/d\) for the inverse subordinator.

\section{Conclusion}\label{conclusion}

We have explored the use of an algorithm which is based on the
Semi-Markov property of CTRW limits. To achieve a concise and general
representation of CTRW limits, we have identified CTRW limits with a
bivariate Langevin process, which in turn is defined via a coefficient
tuple \((a(x,t), b(x,t), d(x), \overline \nu(w|x))\). Given any such
tuple, we can compute probability densities of the CTRW limit at any
given time.

The main novel settings to which our algorithm applies are:

\begin{itemize}
\tightlist
\item
  Spatially varying exponents: we have explored two variants of an
  interface problem, with spatially varying anomalous exponent and
  spatially varying mixture of two anomalous exponents.
\item
  Temporal drift: a drift added to the Levy flight \(Z_u\) translates
  into a ``speed limit'' for the time evolution \(E(t)\), a phenomenon
  which changes the behaviour at short times of CTRW limits and which is
  seemingly unknown in the statistical physics literature.
\item
  Inverse subordinators: these are main building blocks for anomalous
  diffusion problems, and our algorithm computes their densities in
  great generality.
\end{itemize}

Contrary to popular knowledge, Semi-Markov processes are not necessarily
discontinuous piecewise constant processes with state-dependent holding
time distributions. Semi-Markov processes include CTRW limits (with
continuous sample trajectories), an idea which we have exploited in this
paper. They also include \emph{coupled} CTRW limits (Straka and Henry
2011) and, in a wider sense, Levy walks (Magdziarz et al. 2015). The
main idea from this paper, i.e.~leveraging the Semi-Markov property to
compute probability densities, can also be applied to these types of
processes, which we deem an interesting future extension of the present
work.

\subsection*{Acknowledgements}\label{acknowledgements}
\addcontentsline{toc}{subsection}{Acknowledgements}

Peter Straka was supported by the Australian Research Council with a
Discovery Early Career Researcher Award (DECRA) DE160101147. The authors
thank Christopher Angstmann, Bruce Henry and James Nichols for helpful
discussions on discrete time random walks.

\subsection*{Reproducibility}\label{reproducibility}
\addcontentsline{toc}{subsection}{Reproducibility}

All computations and plots of this paper were made using the \texttt{R}
programming language (R Core Team 2018) with the \texttt{rmarkdown}
(Allaire, Xie, McPherson, et al. 2018) and \texttt{rticles} (Allaire,
Xie, R Foundation, et al. 2018) packages. All source code is openly
available (Straka and Gill 2018).

\section*{Appendix}\label{appendix}
\addcontentsline{toc}{section}{Appendix}

\appendix

\section{\texorpdfstring{Checking conditions \eqref{eq:cond1} --
\eqref{eq:cond4}}{Checking conditions  -- }}\label{checking-conditions}

The following lemma pertains to the calculations in the waiting times of
\eqref{eq:cond1} -- \eqref{eq:cond4}:

\begin{lemma} \label{lem:psi}
Under condition \eqref{eq:blowup}, 
the waiting time distribution \eqref{eq:def-psi} satisfies, as $c \to \infty$,
\begin{align}
\label{eq:psi2delta}
\int f(w) \psi^{(c)}(w|y)\,dw = \sum_{j = 1}^\infty f(j\tau) \psi^{(c)}(j\tau|y)
&\to f(0),
\\ \label{eq:psi-non-local}
c \int g(w) \psi^{(c)}(w|y)\,dw
= c \sum\limits_{j\tau > 0} g(j\tau) \psi^{(c)}(j\tau|y)
&\to \int g(w) \nu(w|y)\,dw,
\\ \label{eq:psi-local}
c \int_0^\varepsilon w \psi^{(c)}(w|y)\,dw
= c \sum\limits_{0 < j\tau \le \varepsilon} j\tau \psi^{(c)}(j\tau | y)
&\to d(y)
+ \mathcal O\left(\frac{\varepsilon^{1-\beta(y)}}{\Gamma(1-\beta(y))}\right), \quad \varepsilon > 0.
\end{align}
for any bounded continuous $f$ and $g$, where $g$ vanishes in a 
neighbourhood of $0$.
\end{lemma}

\emph{Proof.} \eqref{eq:psi2delta} holds since \(\psi^{(c)}(w|y)\) is a
probability distribution on the positive numbers with tail function \[
\Psi^{(c)}(w|x) = d(x) \mathbf 1\{w \le \tau\}  + (1-d(x)) H^{(c)}(w_\tau|x)
\] which for all \(w>0\) satisfies \(\Psi^{(c)}(w|y) \to 0\) as
\(c \to \infty\) (recall that \(\tau = 1/c \downarrow 0\)). For
\eqref{eq:psi-non-local}, we first note that

\begin{align} \label{eq:tail-to-zero}
c \Psi^{(c)}(w|y)
= c d(y) \mathbf 1(w \le \tau) 
+ [c(1-d(y))] \wedge \overline \nu(w_\tau) \to \overline \nu(w), 
\quad c \to \infty,
\end{align}

for every \(w > 0\). Assume that \(g\) is differentiable, and let
\(\varepsilon > 0\) be small enough so that \(g(\varepsilon) = 0\).
Using (Lebesgue-Stieltjes) integration by parts, we may calculate

\begin{align*}
c \int_0^\infty g(w) \psi^{(c)}(w|y)\,dw
= c \int_\varepsilon^\infty g(w) \psi^{(c)}(w|y)\,dw
= c \int_\varepsilon^\infty g'(w) \Psi^{(c)}(w|y)\,dw
\\
\to \int_\varepsilon^\infty g'(w) \overline \nu(w|y)\,dw
= \int_\varepsilon^\infty g(w) \nu(w|y)\,dw
= \int_0^\infty g(w) \nu(w|y)\,dw.
\end{align*}

But bounded continuous functions can be approximated by differentiable
functions with arbitrary accuracy, so \eqref{eq:psi-non-local} follows.

Finally, for \eqref{eq:psi-local}, we consider the local and nonlocal
parts \(\psi^{(c)}_{\rm loc}(w|x)\) and \(\psi^{(c)}_{\rm nonloc}(w|x)\)
separately. For the local part, we have \[
c \int_0^\varepsilon w\, \psi^{(c)}_{\rm loc}(w|x)\,dw
= c \tau \to 1.
\] For the nonlocal part, we use Lebesgue-Stieltjes integration by
parts:

\begin{align*}
&\int_0^\varepsilon w\, \psi^{(c)}_{\rm nonloc}(w|x)\,dw
=
\int_0^\varepsilon w\, \left(-d\Psi^{(c)}_{\rm nonloc}(w|x) \right)
\\
&= - \left[ w \Psi^{(c)}_{\rm nonloc}(w|x)\right]_0^\varepsilon
+ \int_0^\varepsilon \Psi^{(c)}_{\rm nonloc}(w|x)\,dw
\\
&= - \varepsilon \Psi^{(c)}_{\rm nonloc}(\varepsilon|x)
+ \int_0^\varepsilon \Psi^{(c)}_{\rm nonloc}(w|x)\,dw
\end{align*}

Multiplying with \(c\) and letting \(c \to \infty\), the right hand side
converges to \[
-\frac{\varepsilon \overline \nu(\varepsilon|x)}{1-d(x)}
+ \int_0^\varepsilon \frac{\overline \nu(\varepsilon|x)}{1-d(x)}\,dw
\] where both terms are of order
\(\mathcal O\left(\varepsilon^{1-\beta(x)} / \Gamma(1-\beta(x))\right)\)
according to the technical assumption \eqref{eq:blowup}.
\eqref{eq:psi-local} now follows from the definition \eqref{eq:def-psi}
of \(\psi^{(c)}(w|x)\). \qed

The final lemma pertains to the jump distributions in
\eqref{eq:cond1}--\eqref{eq:cond4}:

\begin{lemma}\label{lem:jumps}
The jump probabilities $\ell^{(c)}(x,t), r^{(c)}(x,t)$ and 
$n(x,t)$ satisfy
\begin{align}
\label{eq:jump-calc-b}
c [-\chi \ell^{(c)}(x,t+w) + \chi r^{(c)}(x,t+w)]
&= b(x,t+w)
\\ \label{eq:jump-calc-a}
c \chi^2 [\ell^{(c)}(x,t+w) + r^{(c)}(x,t+w)] &= a(x,t+w)
\\ \label{eq:jump2delta}
\int_\mathbb R f(z)\left[r^{(c)}(x,t) \delta_\chi(z)
+ n(x,t) \delta_0(z) + \ell^{(c)}(x,t) \delta_{-\chi}(z) \right]\,dz
&\to f(0)
\end{align}
as $c \to \infty$ for all bounded continuous $f$.
\end{lemma}

\emph{Proof.} This follows easily from the definitions of the jump
probabilities. \qed

Finally, to see that \eqref{eq:cond1}--\eqref{eq:cond2} hold, use
\eqref{eq:jump-calc-b}--\eqref{eq:jump-calc-a} and \eqref{eq:psi2delta}.
To see \eqref{eq:cond3}, use \eqref{eq:psi-local} and let
\(\varepsilon \downarrow 0\); and finally, to see \eqref{eq:cond4}, use
\eqref{eq:psi-non-local} and \eqref{eq:jump2delta}.

\section*{References}\label{references}
\addcontentsline{toc}{section}{References}

\hypertarget{refs}{}
\hypertarget{ref-rmarkdown}{}
Allaire, JJ, Yihui Xie, Jonathan McPherson, Javier Luraschi, Kevin
Ushey, Aron Atkins, Hadley Wickham, Joe Cheng, and Winston Chang. 2018.
\emph{Rmarkdown: Dynamic Documents for R}.
\url{https://CRAN.R-project.org/package=rmarkdown}.

\hypertarget{ref-rticles}{}
Allaire, JJ, Yihui Xie, R Foundation, Hadley Wickham, Journal of
Statistical Software, Ramnath Vaidyanathan, Association for Computing
Machinery, et al. 2018. \emph{Rticles: Article Formats for R Markdown}.
\url{https://CRAN.R-project.org/package=rticles}.

\hypertarget{ref-Alrawashdeh2017}{}
Alrawashdeh, Mahmoud S., James F. Kelly, Mark M Meerschaert, and Hans
Peter Scheffler. 2017. ``Applications of inverse tempered stable
subordinators.'' \emph{Comput. Math. with Appl.} 73 (6). Elsevier Ltd:
892--905.
doi:\href{https://doi.org/10.1016/j.camwa.2016.07.026}{10.1016/j.camwa.2016.07.026}.

\hypertarget{ref-Angstmann2016a}{}
Angstmann, C.N., I.C. Donnelly, Bruce I Henry, B.A. Jacobs, T.A.M.
Langlands, and J.A. Nichols. 2016. ``From stochastic processes to
numerical methods: A new scheme for solving reaction subdiffusion
fractional partial differential equations.'' \emph{J. Comput. Phys.} 307
(February). Elsevier Inc.: 508--34.
doi:\href{https://doi.org/10.1016/j.jcp.2015.11.053}{10.1016/j.jcp.2015.11.053}.

\hypertarget{ref-Angstmann2015a}{}
Angstmann, Christopher N, Isaac C Donnelly, Bruce I Henry, and James A
Nichols. 2015. ``A discrete time random walk model for anomalous
diffusion.'' \emph{J. Comput. Phys.} 293. Elsevier Inc.: 53--69.
doi:\href{https://doi.org/10.1016/j.jcp.2014.08.003}{10.1016/j.jcp.2014.08.003}.

\hypertarget{ref-Angstmann2015}{}
Angstmann, Christopher N, Isaac C Donnelly, Bruce I Henry, T. A. M.
Langlands, and Peter Straka. 2015. ``Generalized Continuous Time Random
Walks, Master Equations, and Fractional Fokker--Planck Equations.''
\emph{SIAM J. Appl. Math.} 75 (4): 1445--68.
doi:\href{https://doi.org/10.1137/15M1011299}{10.1137/15M1011299}.

\hypertarget{ref-Applebaum}{}
Applebaum, D. 2009. \emph{Lévy Processes and Stochastic Calculus}. Book.
2nd ed. Vol. 116. Cambridge Studies in Advanced Mathematics. Cambridge
University Press.

\hypertarget{ref-BaeumerStraka16}{}
Baeumer, Boris, and Peter Straka. 2016. ``Fokker--Planck and Kolmogorov
Backward Equations for Continuous Time Random Walk scaling limits.''
\emph{Proc. Am. Math. Soc.}, 1--14.
doi:\href{https://doi.org/10.1090/proc/13203}{10.1090/proc/13203}.

\hypertarget{ref-Banks2005}{}
Banks, Daniel S., and Cécile Fradin. 2005. ``Anomalous diffusion of
proteins due to molecular crowding.'' \emph{Biophys. J.} 89 (5):
2960--71.
doi:\href{https://doi.org/10.1529/biophysj.104.051078}{10.1529/biophysj.104.051078}.

\hypertarget{ref-Berkowitz2008}{}
Berkowitz, Brian, Simon Emmanuel, and H. Scher. 2008. ``Non-Fickian
transport and multiple-rate mass transfer in porous media.'' \emph{Water
Resour. Res.} 44 (3): 1--16.
doi:\href{https://doi.org/10.1029/2007WR005906}{10.1029/2007WR005906}.

\hypertarget{ref-Chen2010}{}
Chen, Chang-Ming, F. Liu, V. Anh, and I. Turner. 2010. ``Numerical
Schemes with High Spatial Accuracy for a Variable-Order Anomalous
Subdiffusion Equation.'' \emph{SIAM J. Sci. Comput.} 32 (4): 1740--60.
doi:\href{https://doi.org/10.1137/090771715}{10.1137/090771715}.

\hypertarget{ref-Fedotov2012}{}
Fedotov, Sergei, and Steven Falconer. 2012. ``Subdiffusive master
equation with space-dependent anomalous exponent and structural
instability.'' \emph{Phys. Rev. E} 85 (3): 031132.
doi:\href{https://doi.org/10.1103/PhysRevE.85.031132}{10.1103/PhysRevE.85.031132}.

\hypertarget{ref-Gajda2010}{}
Gajda, Janusz, and Marcin Magdziarz. 2010. ``Fractional Fokker-Planck
equation with tempered \(\alpha\)-stable waiting times: Langevin picture
and computer simulation.'' \emph{Phys. Rev. E} 82 (1): 1--6.
doi:\href{https://doi.org/10.1103/PhysRevE.82.011117}{10.1103/PhysRevE.82.011117}.

\hypertarget{ref-gardiner2004handbook}{}
Gardiner, C.W. 2004. \emph{Handbook of Stochastic Methods for Physics,
Chemistry, and the Natural Sciences}. Springer Complexity. Springer.
\url{https://books.google.com.au/books?id=wLm7QgAACAAJ}.

\hypertarget{ref-Gill2016}{}
Gill, Gurtek, and Peter Straka. 2016. ``A Semi-Markov Algorithm for
Continuous Time Random Walk Limit Distributions.'' Edited by A.
Nepomnyashchy and V. Volpert. \emph{Math. Model. Nat. Phenom.} 11 (3):
34--50.
doi:\href{https://doi.org/10.1051/mmnp/201611303}{10.1051/mmnp/201611303}.

\hypertarget{ref-Hahn11}{}
Hahn, Marjorie G, Kei Kobayashi, J. Ryvkina, and Sabir Umarov. 2011.
``On time-changed Gaussian processes and their associated
Fokker-Planck-Kolmogorov equations.'' \emph{Electron. Commun. Probab.}
16: 150--64.
\url{http://www.emis.ams.org/journals/EJP-ECP/_ejpecp/ECP/include/getdocc776.pdf?id=5619\&article=2284\&mode=pdf}.

\hypertarget{ref-Hanert2014}{}
Hanert, Emmanuel, and Cécile Piret. 2014. ``A Chebyshev PseudoSpectral
Method to Solve the Space-Time Tempered Fractional Diffusion Equation.''
\emph{SIAM J. Sci. Comput.} 36 (4): A1797--A1812.
doi:\href{https://doi.org/10.1137/130927292}{10.1137/130927292}.

\hypertarget{ref-HLS10PRL}{}
Henry, Bruce I, T. A. M. Langlands, and Peter Straka. 2010. ``Fractional
Fokker-Planck Equations for Subdiffusion with Space- and Time-Dependent
Forces.'' Journal article. \emph{Phys. Rev. Lett.} 105 (17). American
Physical Society: 170602.
doi:\href{https://doi.org/10.1103/PhysRevLett.105.170602}{10.1103/PhysRevLett.105.170602}.

\hypertarget{ref-HLS2010b}{}
Henry, Bruce I, T. A.M. Langlands, Peter Straka, and T. A. M. Langlands.
2010. ``An introduction to fractional diffusion.'' Journal article. In
\emph{Complex Phys. Biophys. Econophysical Syst. World Sci. Lect. Notes
Complex Syst.}, edited by R L. Dewar and F Detering, 9:37--90. World
Scientific Lecture Notes in Complex Systems. Singapore: World
Scientific.
doi:\href{https://doi.org/10.1142/9789814277327_0002}{10.1142/9789814277327\_0002}.

\hypertarget{ref-Hofling2012}{}
Höfling, Felix, Thomas Franosch, and Review Article. 2012. ``Anomalous
transport in the crowded world of biological cells,'' 1--55.

\hypertarget{ref-Korabel2010}{}
Korabel, Nickolay, and Eli Barkai. 2010. ``Paradoxes of subdiffusive
infiltration in disordered systems.'' \emph{Phys. Rev. Lett.} 104 (17):
1--4.
doi:\href{https://doi.org/10.1103/PhysRevLett.104.170603}{10.1103/PhysRevLett.104.170603}.

\hypertarget{ref-Langlands2005a}{}
Langlands, T. A. M., and Bruce I Henry. 2005. ``The accuracy and
stability of an implicit solution method for the fractional diffusion
equation.'' \emph{J. Comput. Phys.} 205 (2): 719--36.
doi:\href{https://doi.org/10.1016/j.jcp.2004.11.025}{10.1016/j.jcp.2004.11.025}.

\hypertarget{ref-Li2009}{}
Li, Xianjuan, and Chuanju Xu. 2009. ``A Space-Time Spectral Method for
the Time Fractional Diffusion Equation.'' \emph{SIAM J. Numer. Anal.} 47
(3): 2108--31.
doi:\href{https://doi.org/10.1137/080718942}{10.1137/080718942}.

\hypertarget{ref-Magdziarz2015}{}
Magdziarz, Marcin, Hans-Peter Scheffler, Peter Straka, and P.d
Zebrowski. 2015. ``Limit theorems and governing equations for Lévy
walks.'' \emph{Stoch. Process. Their Appl.} 125 (11). Elsevier B.V.:
4021--38.
doi:\href{https://doi.org/10.1016/j.spa.2015.05.014}{10.1016/j.spa.2015.05.014}.

\hypertarget{ref-MMMPS12}{}
Meerschaert, Mark M, and Peter Straka. 2012. ``Fractional Dynamics at
Multiple Times.'' \emph{J. Stat. Phys.} 149 (5): 878--86.
doi:\href{https://doi.org/10.1007/s10955-012-0638-z}{10.1007/s10955-012-0638-z}.

\hypertarget{ref-invSubord}{}
---------. 2013. ``Inverse Stable Subordinators.'' Edited by A.
Nepomnyashchy and V. Volpert. \emph{Math. Model. Nat. Phenom.} 8 (2):
1--16.
doi:\href{https://doi.org/10.1051/mmnp/20138201}{10.1051/mmnp/20138201}.

\hypertarget{ref-Meerschaert2014}{}
---------. 2014. ``Semi-Markov approach to continuous time random walk
limit processes.'' \emph{Ann. Probab.} 42 (4): 1699--1723.
doi:\href{https://doi.org/10.1214/13-AOP905}{10.1214/13-AOP905}.

\hypertarget{ref-Metzler2000}{}
Metzler, Ralf, and Joseph Klafter. 2000. ``The random walk's guide to
anomalous diffusion: a fractional dynamics approach.'' Journal article.
\emph{Phys. Rep.} 339 (1). Elsevier: 1--77.
doi:\href{https://doi.org/10.1016/S0370-1573(00)00070-3}{10.1016/S0370-1573(00)00070-3}.

\hypertarget{ref-Mustapha2011}{}
Mustapha, Kassem, and William McLean. 2011. ``Piecewise-linear,
discontinuous Galerkin method for a fractional diffusion equation.''
\emph{Numer. Algorithms} 56 (2): 159--84.
doi:\href{https://doi.org/10.1007/s11075-010-9379-8}{10.1007/s11075-010-9379-8}.

\hypertarget{ref-Orsingher2018}{}
Orsingher, Enzo, Costantino Ricciuti, and Bruno Toaldo. 2018. ``On
semi-Markov processes and their Kolmogorov's integro-differential
equations.'' \emph{J. Funct. Anal.} 275 (4). Elsevier Inc.: 830--68.
doi:\href{https://doi.org/10.1016/j.jfa.2018.02.011}{10.1016/j.jfa.2018.02.011}.

\hypertarget{ref-R}{}
R Core Team. 2018. \emph{R: A Language and Environment for Statistical
Computing}. Vienna, Austria: R Foundation for Statistical Computing.
\url{https://www.R-project.org/}.

\hypertarget{ref-Regner2013}{}
Regner, Benjamin M., Dejan Vučinić, Cristina Domnisoru, Thomas M.
Bartol, Martin W. Hetzer, Daniel M. Tartakovsky, and Terrence J.
Sejnowski. 2013. ``Anomalous diffusion of single particles in
cytoplasm.'' \emph{Biophys. J.} 104 (8): 1652--60.
doi:\href{https://doi.org/10.1016/j.bpj.2013.01.049}{10.1016/j.bpj.2013.01.049}.

\hypertarget{ref-Sandev2015}{}
Sandev, Trifce, Aleksei V. Chechkin, Nickolay Korabel, Holger Kantz,
Igor M. Sokolov, and Ralf Metzler. 2015. ``Distributed-order diffusion
equations and multifractality: Models and solutions.'' \emph{Phys. Rev.
E - Stat. Nonlinear, Soft Matter Phys.} 92 (4): 1--19.
doi:\href{https://doi.org/10.1103/PhysRevE.92.042117}{10.1103/PhysRevE.92.042117}.

\hypertarget{ref-Santamaria2006a}{}
Santamaria, Fidel, Stefan Wils, Erik De Schutter, and George J.
Augustine. 2006. ``Anomalous diffusion in Purkinje cell dendrites caused
by spines.'' \emph{Neuron} 52 (4): 635--48.
doi:\href{https://doi.org/10.1016/j.neuron.2006.10.025}{10.1016/j.neuron.2006.10.025}.

\hypertarget{ref-Savov2018}{}
Savov, Mladen, and Bruno Toaldo. 2018. ``Semi-Markov processes,
integro-differential equations and anomalous diffusion-aggregation,''
1--37. \url{http://arxiv.org/abs/1807.07060}.

\hypertarget{ref-Scalas2006}{}
Scalas, Enrico. 2006. ``The application of continuous-time random walks
in finance and economics.'' \emph{Phys. A Stat. Mech. Its Appl.} 362:
225--39.

\hypertarget{ref-SchumerMIM}{}
Schumer, Rina, David A Benson, Mark M Meerschaert, and Boris Baeumer.
2003. ``Fractal mobile/immobile solute transport.'' \emph{Water Resour.
Res.} 39 (10).
doi:\href{https://doi.org/10.1029/2003WR002141}{10.1029/2003WR002141}.

\hypertarget{ref-Sokolov2006e}{}
Sokolov, Igor M, and Joseph Klafter. 2006. ``Field-Induced Dispersion in
Subdiffusion.'' \emph{Phys. Rev. Lett.} 97 (14): 1--4.
doi:\href{https://doi.org/10.1103/PhysRevLett.97.140602}{10.1103/PhysRevLett.97.140602}.

\hypertarget{ref-Stickler2011}{}
Stickler, B. A., and E. Schachinger. 2011. ``Continuous time anomalous
diffusion in a composite medium.'' \emph{Phys. Rev. E - Stat. Nonlinear,
Soft Matter Phys.} 84 (2): 1--9.
doi:\href{https://doi.org/10.1103/PhysRevE.84.021116}{10.1103/PhysRevE.84.021116}.

\hypertarget{ref-StrakaThesis}{}
Straka, Peter. 2011. ``Continuous Time Random Walk Limit Processes:
Stochastic Models for Anomalous Diffusion.'' PhD thesis, University of
New South Wales.
\url{http://unsworks.unsw.edu.au/fapi/datastream/unsworks:9800/SOURCE02}.

\hypertarget{ref-Straka17}{}
---------. 2018. ``Variable order fractional Fokker--Planck equations
derived from Continuous Time Random Walks.'' \emph{Physica A:
Statistical Mechanics and Its Applications} 503 (August): 451--63.
doi:\href{https://doi.org/10.1016/j.physa.2018.03.010}{10.1016/j.physa.2018.03.010}.

\hypertarget{ref-StrakaFedotov14}{}
Straka, Peter, and Sergei Fedotov. 2015. ``Transport equations for
subdiffusion with nonlinear particle interaction.'' \emph{J. Theor.
Biol.} 366 (February). Elsevier: 71--83.
doi:\href{https://doi.org/10.1016/j.jtbi.2014.11.012}{10.1016/j.jtbi.2014.11.012}.

\hypertarget{ref-varyExp}{}
Straka, Peter, and Gurtek Gill. 2018. ``Strakaps/Varyexp V1.0.''
doi:\href{https://doi.org/10.5281/zenodo.1346346}{10.5281/zenodo.1346346}.

\hypertarget{ref-StrakaHenry}{}
Straka, Peter, and Bruce I Henry. 2011. ``Lagging and leading coupled
continuous time random walks, renewal times and their joint limits.''
\emph{Stoch. Process. Their Appl.} 121 (2). Elsevier B.V.: 324--36.
doi:\href{https://doi.org/10.1016/j.spa.2010.10.003}{10.1016/j.spa.2010.10.003}.

\hypertarget{ref-TMT04}{}
Tolić-Nørrelykke, Iva Marija, Emilia-Laura Munteanu, Genevieve Thon,
Lene Oddershede, and Kirstine Berg-Sørensen. 2004. ``Anomalous Diffusion
in Living Yeast Cells.'' \emph{Phys. Rev. Lett.} 93 (7): 078102.
doi:\href{https://doi.org/10.1103/PhysRevLett.93.078102}{10.1103/PhysRevLett.93.078102}.

\hypertarget{ref-Weron2008}{}
Weron, A., and Marcin Magdziarz. 2008. ``Modeling of subdiffusion in
space-time-dependent force fields beyond the fractional Fokker-Planck
equation.'' Journal article. \emph{Phys. Rev. E} 77 (3). APS: 1--6.
doi:\href{https://doi.org/10.1103/PhysRevE.77.036704}{10.1103/PhysRevE.77.036704}.

\hypertarget{ref-Whitt2010}{}
Whitt, Ward. 2001. \emph{Stochastic-Process Limits: An Introduction to
Stochastic-Process Limits and their Application to Queues}. Book. 1st
ed. New York: Springer.

\hypertarget{ref-Wong04}{}
Wong, I Y, M L Gardel, D R Reichman, Eric R Weeks, M T Valentine, A R
Bausch, and D A Weitz. 2004. ``Anomalous Diffusion Probes Microstructure
Dynamics of Entangled F-Actin Networks.'' \emph{Phys. Rev. Lett.} 92
(17): 178101.
doi:\href{https://doi.org/10.1103/PhysRevLett.92.178101}{10.1103/PhysRevLett.92.178101}.

\hypertarget{ref-Yuste2005}{}
Yuste, S B, and L Acedo. 2005. ``An Explicit Finite Difference Method
and a New von Neumann-Type Stability Analysis for Fractional Diffusion
Equations'' 42 (5): 1862--74.
doi:\href{https://doi.org/10.1137/030602666}{10.1137/030602666}.

\end{document}